\documentclass[letterpaper,twocolumn,showpacs,preprintnumbers,amsmath,amssymb]{revtex4}
\usepackage{graphicx}

\begin{document}

\preprint{Published as: J.~A. Petrus, P. Bohlouli-Zanjani, J.~D.~D. Martin,
J.~Phys.~B.~At.~Mol.~Opt.~Phys., v.~41, 245001 (2008).} 

\title[ac electric-field-induced resonant energy transfer]{
ac electric-field-induced resonant energy transfer 
between cold Rydberg atoms}

\author{J.~A. Petrus, P. Bohlouli-Zanjani, J.~D.~D. Martin}

\address{Department of Physics and Astronomy, 
University of Waterloo, Waterloo, ON, \mbox{N2L 3G1}, Canada}

\begin{abstract}
An oscillating electric field at 1.356 GHz
was used to promote the resonant energy transfer process: 
$43d_{5/2}+43d_{5/2} \rightarrow 45p_{3/2}+41f$ 
between translationally cold $^{85}$Rb Rydberg atoms.  
The ac Stark shifts due to this dressing field created degeneracies
between the initial and final two-atom states of this process.
The ac field strength was scanned to collect spectra which are 
analogous to dc electric-field-induced resonant energy transfer spectra.  
Different resonances were observed for different magnetic sublevels 
involved in the process. Compared to earlier work performed at higher 
frequencies, the choice of dressing frequency and structure of the spectra 
may be intuitively understood, by analogy with the dc field case.
\end{abstract}


\maketitle

\section{Introduction}

The interaction energies between neighbouring Rydberg atoms can be much
larger than between ground state atoms, separated by the same distance.
These interactions can be made even larger if they are resonant.
For example, Vogt {\it et al.}~\cite{vogt:2006} have 
demonstrated that optical
transitions to Cs Rydberg states are partially inhibited by
tuning the energy of two Cs atoms in the $38p_{3/2}$ 
state to be identical to the energy of a $38s_{1/2}$, $39s_{1/2}$
atom pair.  This tuning was accomplished using small electric fields 
($\approx 1-2 \: {\rm V/cm}$).  With this induced degeneracy, the
electric dipole-dipole interaction gives a first order energy shift,
rather than a weaker second-order effect (van der Waals).

The use of external fields to create resonant interactions is common
in atomic physics.  Tunable dc magnetic fields may be used to create
Feshbach resonances, dramatically enhancing the interactions between
ultracold atoms \cite{inouye:1998,courteille:1998}.
Oscillating fields may also be used.  For example,
Gerbier {\it et al.}~\cite{gerbier:2006} 
recently demonstrated that oscillating
magnetic fields may also be used to shift dressed state energy levels
into resonance, and thereby enhance interatomic interactions between
ultracold atoms.  The 
ability to vary both the frequency and amplitude of the ``dressing
field'' gives additional latitude in achieving the resonance condition.

In a similar vein, we have recently shown that the resonant 
energy transfer process:
\begin{equation}
43d_{5/2}+43d_{5/2} \rightarrow 45p_{3/2}+41f
\label{eq:process}
\end{equation}
in $^{85}{\rm Rb}$ may be greatly 
enhanced by the application of a microwave field (28.5 GHz)
of appropriate field strengths \cite{bohlouli:2007}.  
The results agreed with the
theoretical predictions for the ac Stark effect,
but it is not obvious why 28.5 GHz was an appropriate
frequency to use.  Since all of the participating levels 
($43d_{5/2}$, $45p_{3/2}$ and $41f$) 
show significant shifts at this frequency, the situation
is complicated.  In the present
work, we demonstrate that a much lower frequency of 1.356 GHz
-- chosen to be slightly blue detuned from the $41f-41g$ transition -- shifts 
only $41f$ significantly.  
By varying the dressing field amplitude we can 
shift the process in Eq.~\ref{eq:process} into resonance.  

There is an additional benefit to using a lower dressing frequency.
As discussed
in Ref.~\cite{bohlouli:2007}, the different magnetic sublevels involved
in Eq.~\ref{eq:process} show different ac Stark shifts.
With a 28.5 GHz dressing field,  these give a complicated series of 
resonances. In the present work -- at much lower frequencies --
the spectra are not as complicated, as
the only important magnetic sublevel structure is due to the $41f$
states.

\section{Theoretical Background}

This section begins by motivating the choice of a dressing frequency
of $1.356 \: {\rm GHz}$, and concludes by discussing the techniques used for 
calculating the dressed atom energy levels.

The process given in Eq.~\ref{eq:process} is not resonant in the 
absence of applied fields.  The final state is lower in energy
than the initial state by 
approximately $10 \: {\rm MHz}$ \cite{bohlouli:2007}.
The $41f$ states shift to lower energy with increasing dc electric field.
The other states involved have significantly smaller dc polarizabilities,
and do not shift as much.  Therefore, the resonant energy transfer
process in Eq.~\ref{eq:process} cannot be shifted into resonance
with a dc field.

\begin{figure}
\centerline{\includegraphics[width=8cm]{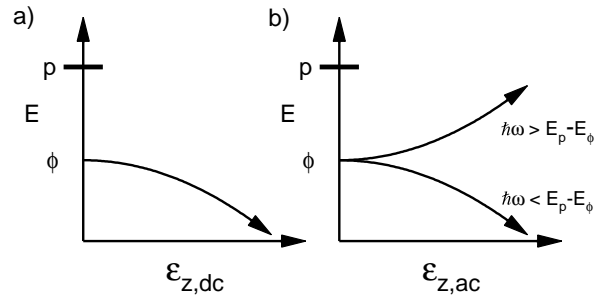}}
\caption{\label{fg:simple}
(a) The dc Stark effect in the case of a single
nearby perturber $|p\!\!>$.  The energy of the perturbed state 
$|\phi\!\!>$ is shown as a function of the dc electric-field magnitude
$\varepsilon_{z,dc}$.
(b) The ac Stark effect in the case of a single nearby perturber 
$|p\!\!>$.  In this case the shifts are shown as a function of the
amplitude of the oscillating electric field $\varepsilon_{z,ac}$.
These two figures illustrate that the ac Stark shift
direction may be altered by an appropriate choice of frequency $\omega$, 
whereas the dc Stark shift direction is fixed. In both cases the perturber
also shifts with field strength; however, this has not been shown.
}
\end{figure}

The shift 
of the $41f$ states to lower energies can be intuitively
understood using 2$^{\rm nd}$ order perturbation theory. 
For an arbitrary state $|\phi\!\!>$, and an electric field
of magnitude $\varepsilon_{z,dc}$ pointing in the $z$ direction:
\begin{equation}
\Delta E_{\phi} = \varepsilon_{z,dc}^2 
\sum_{p \ne \phi} 
\frac{|\!\!<\!\! \phi | \mu_z | p \!\!>\!\!|^2}{(E_{\phi}-E_p)},
\label{eq:dcshift}
\end{equation}
where $E_p$ refers to the energy of state $|p\!\!>$ and
$\mu_z$ is the electric dipole moment in the $z$ direction.
Figure \ref{fg:simple}a illustrates the simplified case of one dominant
perturber.  States are pushed away in energy from the perturber.
For the $41f$ state, the dominant nearby perturber is the $41g$ state,
located 1.19 GHz higher in energy.  As an electric field is applied,
the $41f$ state shifts down in energy, away from $41g$.

The expression for the perturbative ac Stark shift is similar 
to that for the dc case:
\begin{equation}
\Delta E_{\phi} = \frac{1}{2} \varepsilon_{z,ac}^2 
\sum_{p \ne \phi} \frac{(E_{\phi}-E_p) 
|\!\!<\!\! \phi | \mu_z | p \!\!>\!\!|^2}
{(E_{\phi}-E_p)^2 - (\hbar \omega)^2},
\label{eq:acshift}
\end{equation}
where $\omega$ is the angular frequency of a dressing field, pointing
in the $z$ direction with electric field amplitude $\varepsilon_{z,ac}$.
(For a derivation, see Ref.~\cite{haas:2006}.)
Again, Fig.~\ref{fg:simple}b illustrates the simplified case of 
one dominant perturber.  In contrast to the dc effect, $\omega$
may be used to control the sign of the shift.  In our specific
case of Rb, we expect that applying a dressing frequency greater
than the $41f-41g$ transition frequency would cause the $41f$ state
to move up in energy with increasing dressing field strength.
As we demonstrate experimentally in the Apparatus and
Results section, this shifts
the initial and final states of Eq.~\ref{eq:process}
into resonance.

Although the perturbative formula (Eq.~\ref{eq:acshift})
for the ac Stark shift provides insight, some caution is required
in its application.  In particular, the energy shifts of the
$41f$ states that are required to make the process in 
Eq.~\ref{eq:process} degenerate are of the same order of magnitude
as the fine structure splitting between $41f_{5/2}$ and $41f_{7/2}$
levels (2.3 MHz).  
This was not as significant with
a dressing field frequency of 28.5 GHz
as the shifts of the $41f$ states were not as large (the 
$43d_{5/2}$ and $45p_{3/2}$ states shifted more to make up the energy
difference in Eq.~\ref{eq:process}, and their fine structure splitting
is much larger).
To treat this complication 
the dressed energy levels are calculated using a 
Floquet approach \cite{shirley:1965}.
This approach takes the ``bare'' basis states of energy $E_{bi}$
and adds ``sideband'' states of energy $E_{bi}+k \hbar \omega$,
where $k$ is an integer.  The ac field introduces coupling between
states differing in sideband order $k$ by one.  Our basis set and
couplings are very similar to those used in dc electric-field
Stark map calculations \cite{zimmerman:1979}.  These calculations
use the experimentally determined $^{85}{\rm Rb}$ Rydberg energy levels 
\cite{li:2003,han:2006} as input.
A nice description of the Floquet approach  is given in Ref.~\cite{water:1990}.

\begin{figure}
\centerline{\includegraphics[width=8cm]{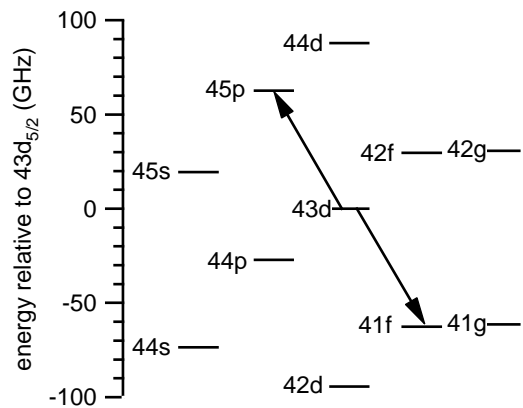}}
\caption{
Some relevant $^{85}$Rb Rydberg atom energy levels.  Fine structure is too
small to be observed on this scale (e.g.~$43d_{3/2}$ and $43d_{5/2}$).
In addition, although the $g$ and $f$ series
appear degenerate, $41g$ is approximately 1.19 GHz higher in energy
than $41f$.  The arrows indicate the transitions involved in the 
ac-electric-field assisted resonant energy transfer 
(see Eq.~\ref{eq:process}).
}
\end{figure}

\section{Apparatus and Results}

\begin{figure}

\centerline{
\includegraphics[width=8cm]{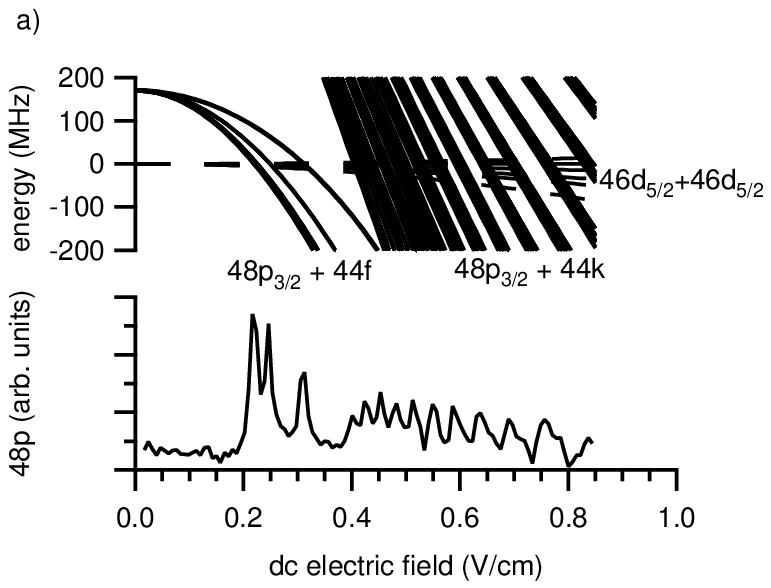}
}

\centerline{
\includegraphics[width=8cm]{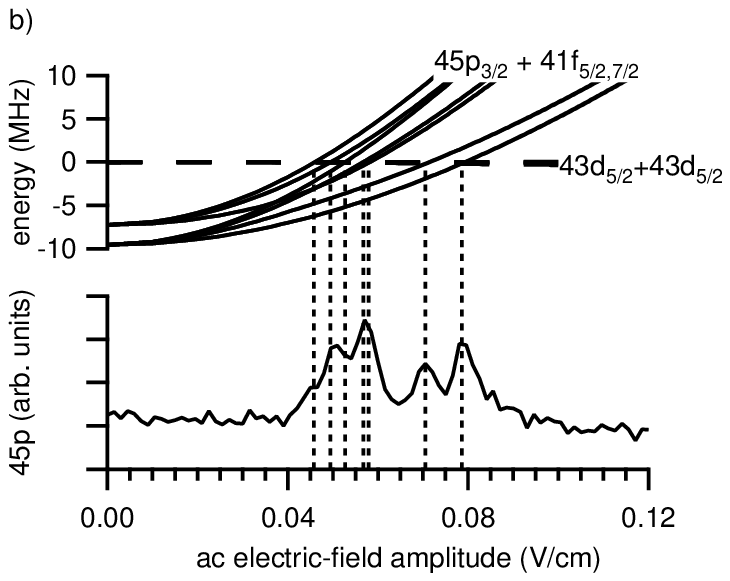}
}

\caption{
\label{fg:retspectra}
(a) dc electric-field-induced resonant energy transfer:
$46d_{5/2}+46d_{5/2} \rightarrow 48p_{3/2}+44f,44k$.  The $44k$
label refers to Stark states -- in this case superpositions of
angular momentum states with $\ell \ge 4$.
(b) ac electric-field-induced resonant energy transfer:
$43d_{5/2}+43d_{5/2} \rightarrow 45p_{3/2}+41f$, using a
dressing frequency of 1.356 GHz.
Note the different horizontal axis scalings for (a) and (b).
Also shown are calculated total energies of initial and final
state atom pairs (for different magnetic sublevel possibilities).
The calculation procedures are discussed in the text.  Vertical
dashed lines are shown in (b) to highlight the agreement between
the calculated and experimentally observed resonance field amplitudes.
}
\end{figure}

We start with a general overview of the experiment.
To observe resonant energy transfer, 
we optically excite cold $^{85}$Rb atoms from a
magneto-optical trap (MOT)
to Rydberg states using a pulse of light.
Then a dc or ac electric field of variable amplitude is applied.  
The Rydberg atoms are allowed to interact in this field
for a fixed amount of time,
and then a selective field ionization pulse is applied to measure the
Rydberg state populations.  If, for example,  
$43d_{5/2}$ Rydberg states are excited
and the ac field establishes the
resonance condition between the initial and final
states shown in Eq.~\ref{eq:process}, then both
$45p_{3/2}$ and $41f$ Rydberg atoms are observed.
By scanning the field strength, repeating the excitation and
detection process, and recording the fraction of Rydberg atoms transferred
to the $45p_{3/2}$ level, 
resonant energy transfer spectra may be obtained
(Fig.~\ref{fg:retspectra}b).

We will now discuss the various aspects of the experiment in more
detail.
The Rydberg atoms are excited using a two-colour scheme involving 
the 780 nm light used for cooling and trapping, and additional
light at approximately 480 nm.  This blue light is generated
by frequency doubling a cw Ti:sapphire laser.  It is introduced
to the atoms in $1-2 \: {\rm \mu s}$ pulses at a repetition rate 
of 10 Hz, using an acousto-optic modulator.  Our frequency stabilization
scheme for the Ti:sapphire laser has been discussed previously
\cite{bohlouli:2006}.

Approximately $25 \: {\rm ms}$ prior to photoexcitation, 
the coil current generating the inhomogeneous
magnetic field necessary for operation of the MOT is switched off.
A $25 \: {\rm ms}$ delay allows the fields due to eddy currents to decay.
By using microwave spectroscopy of a magnetic-field-sensitive 
transition \cite{afrousheh:2006}, 
we have verified that 
the residual magnetic field is less than 0.02 G
at the time of photoexcitation.
Following excitation,
two parallel metal plates surrounding the atoms are used to apply
both dc and ac fields.  One-photon transitions
may also be used to zero-out any residual electric field, and calibrate
applied electric fields \cite{afrousheh2:2006}
(as required for dc electric field assisted
resonant energy transfer spectra -- see Fig.~\ref{fg:retspectra}a). 

Photoexcitation to Rydberg states takes place with no deliberately
applied dc electric field.  For dc-electric-field induced resonant 
energy transfer spectra, the field is ramped up over 160 ns,
immediately following Rydberg excitation (ramping this up relatively
slowly avoids ringing).
In the case of ac-electric-field
induced resonant energy transfer, the oscillating field is switched
on approximately $3 \: {\rm \mu s}$ 
following photoexcitation 
(a delay of this magnitude was not necessary, but it is
important that the dressing field and laser excitation do not 
temporally overlap). 
The Agilent E8254A synthesizer providing this field has a specified rise time
of $100 \: {\rm ns}$.
These field conditions are maintained for approximately
$20 {\rm \: \mu s}$, 
in which time the cold atoms can possibily change
states due to the resonant energy transfer process
(at our densities this time gives maximum amounts of
transfer of about $10-30\%$).
It is essential to switch off the ac field
prior to selective field ionization.
Otherwise, this normally
non-resonant field is shifted into resonance with various transitions
as the electric field increases, creating population transfer unrelated
to interatomic interactions.

Selective field ionization allows the $nd_{5/2}$, $(n+2)p_{3/2}$
and $(n-2)f$ populations
to be distinguished.  All three signals are measured
simultaneously using boxcar integrators, digitized and recorded
as a function of dc or ac field strength.

To illustrate the similarities between ac- and dc-electric-field
induced 
resonant energy transfer spectra, both have been collected
(see Fig.~\ref{fg:retspectra}).  
For the process:
\begin{equation}
\label{eq:generic}
nd_{5/2}+nd_{5/2} \rightarrow (n+2)p_{3/2} + (n-2)f
\end{equation}
the final state is higher in energy than the initial state
for $n \ge 44$, allowing this process to be shifted into
resonance with a dc electric field.   In
Ref.~\cite{bohlouli:2007} an $n=44$ spectrum was presented.  Here 
in Fig.~\ref{fg:retspectra}a we show the case of $n=46$.  
There has been a previous report of this resonant energy transfer
spectrum in the literature \cite{reetz:2006}, but it differs 
from what we have observed, possibly due to differences in signal
to noise.  Note the excellent
agreement between the calculated and observed resonance fields.

As discussed in the introduction, and shown experimentally in
Ref.~\cite{bohlouli:2007}, dc fields cannot be used to shift
the process in Eq.~\ref{eq:generic} into resonance at $n=43$.
However, as illustrated in Fig.~\ref{fg:retspectra}b,
an ac electric field at frequency of 1.356 GHz, approximately 160 MHz higher
than the $41f-41g$ transition, raises
the energy of the final state of Eq.~\ref{eq:generic}, allowing
the resonance condition to be achieved.  The choice of 1.356 GHz
is somewhat arbitrary.  The frequency
of the ac field should be far enough off resonance to 
avoid population transfer, but not so far that unreasonably strong
field amplitudes are required.

Due to unknown impedance mismatchs, we
cannot readily determine the applied ac field strength electrically. 
To calibrate the applied ac electric field amplitude in 
Fig.~\ref{fg:retspectra}b we observed microwave spectra of the
single-photon $43d_{5/2}$ to $41f_{5/2,7/2}$ transitions 
in the presence of a 1.356 GHz dressing field.  These were done at low density
to avoid interatomic effects.  A comparison of observed and
expected shifts (calculated using the Floquet approach) allowed us to assign
field strengths to synthesizer power levels.  As we also use this
calculation to predict the resonance field strengths, we cannot attach
much significance to the overall absolute agreement between the observed
and calculated resonance fields in Fig.~\ref{fg:retspectra}b.  However,
the simultaneous agreement of several resonance peaks 
(corresponding to different magnetic sub-level possibilities) 
suggests that the calculation is correct, as we only used the shift
of the $m_j=1/2$ states for the field strength calibration.

Some practical advantages of applying a relatively low 
frequency field (1.356 GHz here {\it vs.}~28.5 GHz in 
Ref.~\cite{bohlouli:2007}) 
are that the resulting field homogeneity can be much larger due
to the longer corresponding radiation wavelength, and that sources
of lower RF frequencies are more readily available.

\section{Discussion}

We have shown that ac electric fields may be used to induce resonant energy
transfer in a conceptually similar manner as dc electric fields.
This method for enhancing the interactions between Rydberg atoms
may be useful for
``dipole-blockade'' \cite{lukin:2001}.  Local blockade has been
achieved through use of the van der Waals
interaction between Rydberg atoms 
\cite{tong:2004,liebisch:2005} (See also Ref.~\cite{johnson:2008}
for progress on very small numbers of interacting atoms.)
However, it is desirable to
enhance the interatomic interactions by making them resonant.
This has recently been achieved using dc electric
fields \cite{vogt:2006}. The ac Stark shifts induced by 
dressing fields offer an additional means to accomplish this with possible
distinct advantages.  For example in a glass cell (coated with alkali)
it is typically difficult to establish dc electric fields with external 
electrodes, whereas 
high frequency ac fields can be applied more easily (see, for example,
Ref.~\cite{mohapatra:2007}).

Atomic energy levels may be determined by 
dc field induced resonant energy transfer spectra 
(see, for example, Ref.~\cite{afrousheh2:2006}).
The ac field induced
resonant energy transfer process may also be useful 
in this context.  In this and other applications
the frequency of the applied fields allows a variability in the signs
and magnitudes of the Stark shifts involved that gives more flexibility
than the dc shift alone.

This work was supported by NSERC (Canada), CFI, and OIT.

\bibliographystyle{unsrt}
\bibliography{references}

\end{document}